# Physics Archives

## Theories on Phylogenetic Reconstruction

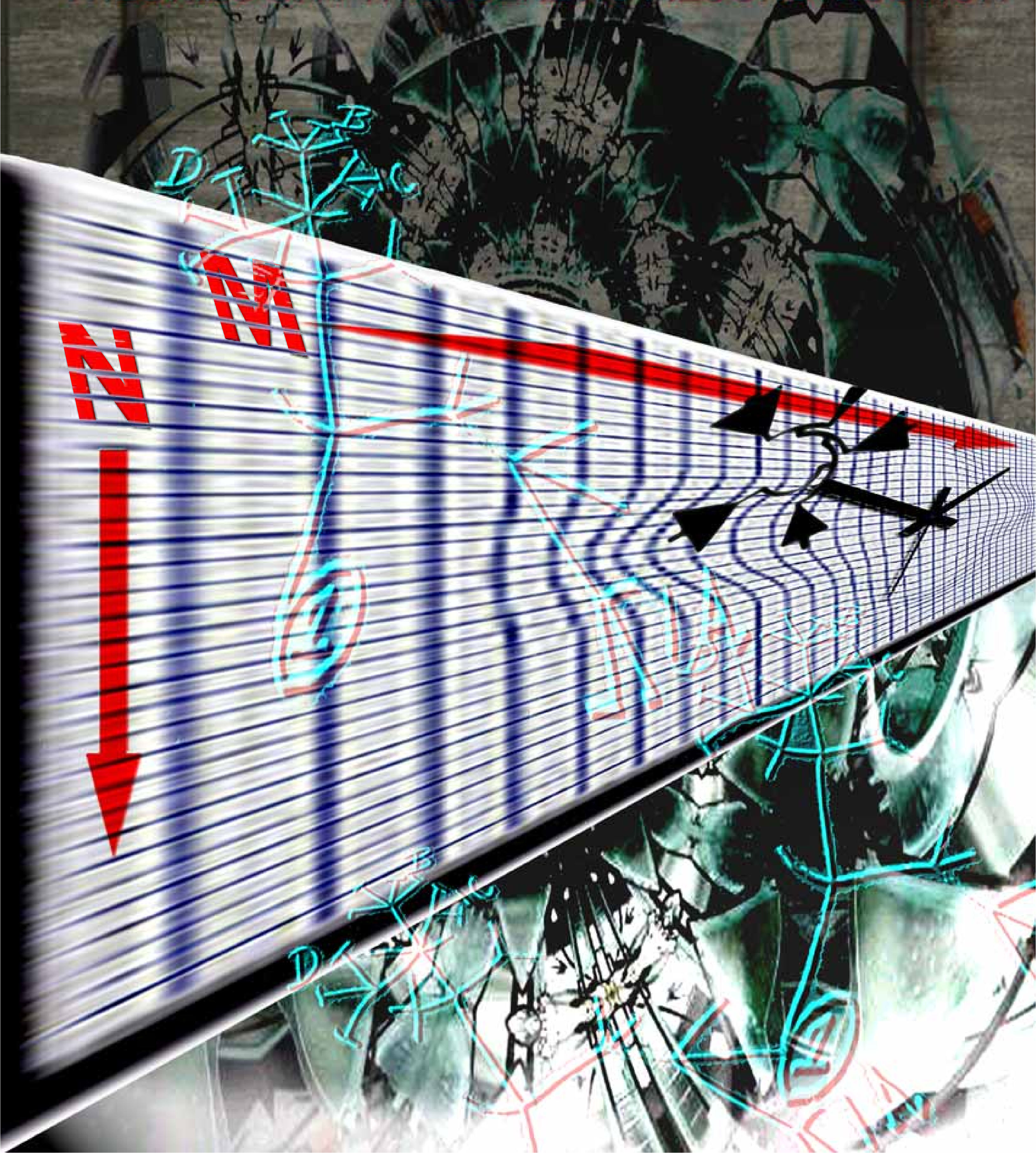

# Theories on PHYlogenetic ReconstructioN (PHYRN)

**Gaurav Bhardwaj[1,2], Zhenhai Zhang[1,3], Yoojin Hong[1,4], Kyung Dae Ko[1,2], Gue Su Chang[3], Evan J. Smith[1,2], Lindsay A. Kline[1,2], D. Nicholas Hartranft[1,2], Edward C. Holmes[1,2], Randen L. Patterson[1,2], and Damian B. van Rossum[1,2].**

(1) Center for Computational Proteomics, The Pennsylvania State University, USA
(2) Department of Biology, The Pennsylvania State University, USA
(3) Department of Biochemistry and Molecular Biology, The Pennsylvania State University, USA
(4) Department of Computer Science and Engineering, The Pennsylvania State University, USA

***Address correspondence to:**

Randen L. Patterson, 230 Life Science Bldg, University Park, PA 16802. Tel: 001-814-865-1668; Fax: 001-814-863-1357; E-mail: rlp25@psu.edu.

Damian B. van Rossum, 518 Wartik Laboratory, University Park, PA 16802. Tel: 001-814-863-1007; Fax: 001-814-863-1357; E-mail: dbv10@psu.edu.

**Abstract**

**The inability to resolve deep node relationships of highly divergent/rapidly evolving protein families is a major factor that stymies evolutionary studies. In this manuscript, we propose a Multiple Sequence Alignment (MSA) independent method to infer evolutionary relationships. We previously demonstrated that phylogenetic profiles built using position specific scoring matrices (PSSMs) are capable of constructing informative evolutionary histories(1;2). In this manuscript, we theorize that PSSMs derived specifically from the query sequences used to construct the phylogenetic tree will improve this method for the study of rapidly evolving proteins. To test this theory, we performed phylogenetic analyses of a benchmark protein superfamily (reverse transcriptases (RT)) as well as simulated datasets. When we compare the results obtained from our method, PHYlogenetic ReconstructioN (PHYRN), with other MSA dependent methods, we observe that PHYRN provides a 4- to 100-fold increase in accurate measurements at deep nodes. As phylogenetic profiles are used as the information source, rather than MSA, we propose PHYRN as a paradigm shift in studying evolution when MSA approaches fail. Perhaps most importantly, due to the improvements in our computational approach and the availability of vast amount of sequencing data, PHYRN is scalable to thousands of sequences. Taken together with PHYRN's adaptability to any protein family, this method can serve as a tool for resolving ambiguities in evolutionary studies of rapidly evolving/highly divergent protein families.**

## Introduction

Phylogenetic profiles have been suggested by us and others as a unified framework for measuring structural, functional, and evolutionary characteristics of protein/protein-families(1;3-6). Proteins within a phylogenetic profile can be defined in an N(query) by M(PSSM) matrix. Under this paradigm, a protein is defined as a vector where each entry quantifies the alignments of a query sequence with a PSSM(3;7). In the case of evolutionary measurements, we previously demonstrated that phylogenetic profiles built in this manner can be used to construct phylogenetic trees using Euclidian distance measurements(1;2).

Our previous study used reverse transcriptases (RT) as a benchmark dataset, and demonstrates that phylogenetic profiles perform well even at extreme levels of divergence (i.e. “twilight zone of sequence similarity”)(2). This work highlighted how pre-existing PSSMs, obtained from the Conserved Domain Database (CDD,(8)), could be utilized to construct an informative M-dimension. When we generated trees with the entire CDD we obtained a tree with perfect monophyly. Despite the perfect monophyly, the statistical support at most deep nodes was lacking. Interestingly, when we analyzed the alignments from the phylogenetic profiles, we determined that the most frequently occurring PSSM alignments were the 16 RT domain-containing profiles present in CDD. When trees were constructed using only these 16RT PSSMs for our M-dimension, we still observed significant monophyly that is well above random (Supplemental Figure 3 in (2)).

Based on these results, it is reasonable to consider that expanding only the informative profiles within our knowledge base will improve the robustness of phylogenetic profile-based measurements, in addition to improving computational performance. In this manuscript, we present data supporting this supposition. Further, we present a pipeline for enriching and amplifying informative PSSMs as well as an algorithmic improvement that drastically reduces computational expense. As evidence for these theories we analyzed biological (RTs) and ROSE-simulated (9) datasets. When compared with other *ab-initio* multiple-sequence alignment (MSA) methods, PHYRN reliably recapitulates true evolutionary history in simulated datasets, and provides deep-node measurements with robust statistical support.

## Results

### PHYRN Pipeline

The algorithm begins by compiling a set of protein queries belonging to the same protein family/superfamily (Figure 1). We use CDD (8) and other approaches to define conserved domains present in members of this superfamily (e.g. RT domain in reverse transcriptases). From this subset of knowledgebase PSSMs we utilize pairwise comparisons to define boundaries of homology. These homologous protein fragments are then utilized to construct a database/library of query-based PSSMs using PSI-BLAST (6-iterations, e-value threshold= $10^{-6}$)(10). In this manner, a query set of 100 sequences can make a library of at least 100 PSSMs. We then use rpsBLAST (8) to obtain pairwise alignments between full-length queries and the query-specific PSSM library. This alignment information (% identity, % coverage) is then encoded into phylogenetic profile matrix. Following, we calculate the Euclidian distance between each query(2). The results from these calculations can then plotted as a phylogenetic tree using a variety of tree-building algorithms (e.g. Neighbor-Joining(11), Maximum Likelyhood(12), Minimum Evolution(13), etc, see Methods for complete description of PHYRN). In the following sections we will highlight the methods for creating informative PSSM libraries and the scoring schemes utilized.

*Enriching and Amplifying Informative PSSMs*

Although CDD provides a comprehensive resource for conserved domains, in all cases the number of PSSMs for any given domain is relatively small (>100). We have previously demonstrated that quality of phylogenetic profile measurements is proportional to the size and variety of the domain-specific PSSM library (2;3;7). This increase in information content is exemplified in Figure 2. When the silkworm Non-LTR AAA92147.1 is analyzed with NCBI CDD, only one RT specific PSSM alignment is returned (Figure 2a). When the same query is analyzed using PHYRN and the 16 RT PSSMs from CDD, we observe 3 overlapping alignments from 3 different PSSMs (~19% of the library, Figure 2b). These overlapping alignments define the boundary from which to generate PSSMs using PSI-BLAST. This same approach was used for 100 RT query sequences; post-expansion, we obtain 102 RT-specific PSSMs. When we reanalyzed the silkworm sequence with the amplified library, ~56% of the PSSM library returns a result within the homologous region.

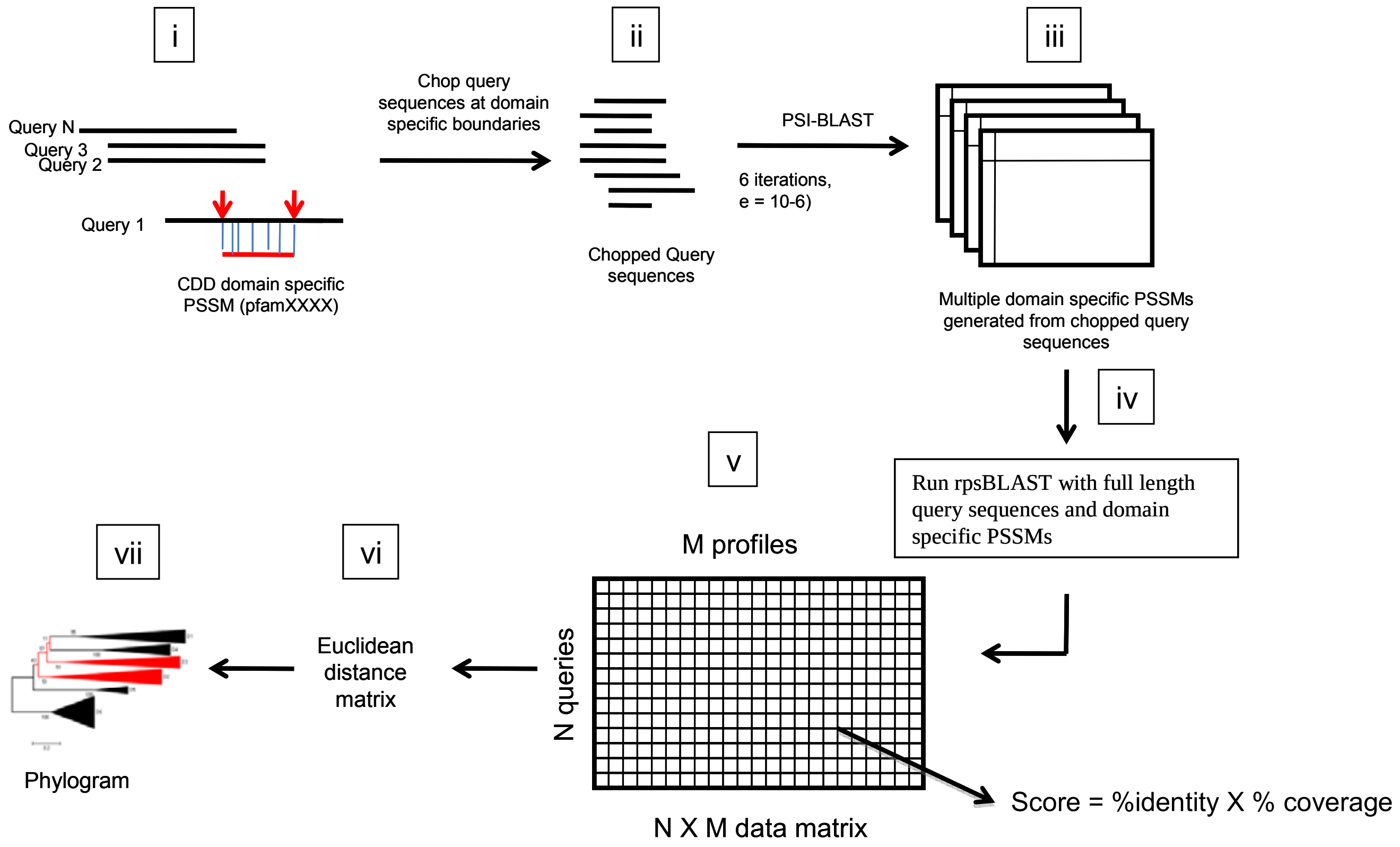

**Figure 1: PHYRN Concept and Work Flow-** PHYRN begins by (i-ii) defining and extracting the domain specific region among the query sequences. (iii) Domain specific regions are then used to create PSSM library using PSI-BLAST. (iv-v) Positive alignments are then calculated between queries and PSSM library using rpsBLAST, and encoded as a PHYRN product score (%identity X %coverage) matrix. (vi) Product score matrix is converted to a Euclidean distance matrix by calculating Euclidean distance between each query pair. vii) Phylogenetic trees can then be graphed using Neighbor-Joining (NJ) or Minimum Evolution (ME) as available in MEGA.

*Scoring Scheme*

In order to encode alignment information between queries and our PSSM libraries, we utilize a product score (%identity X %coverage) during our PHYRN analysis. Equation sets for %identity and %coverage are as defined in (2). The algebraic derivation in Figure 3a demonstrates that our product score is equivalent to (1-p-distance) X gap weight. In data not shown, we determined that the gap weight is a negligible variable, and when removed does not alter our results. Figure 3b depicts the distribution of % gaps, % identity, and % coverage of alignments between 100 RT queries and 102 RT-specific PSSMs. Overall, 92% of alignments are less than 25% identity and the average percentage identity between 100 RT sequences is 21.8% (± 5.6% s.d.). Within a smaller subset of 88 RT sequences, 3644 pairs (95.2%) among 3828 possible pairs of these 88 sequences have less than 25% sequence identity. As a whole RT sequences reside in the "twilight zone" of sequence similarity, underscoring the reason why deducing evolutionary relationships within the RT family is extremely challenging. Furthermore, % gap and % coverage measurements have wide variances.

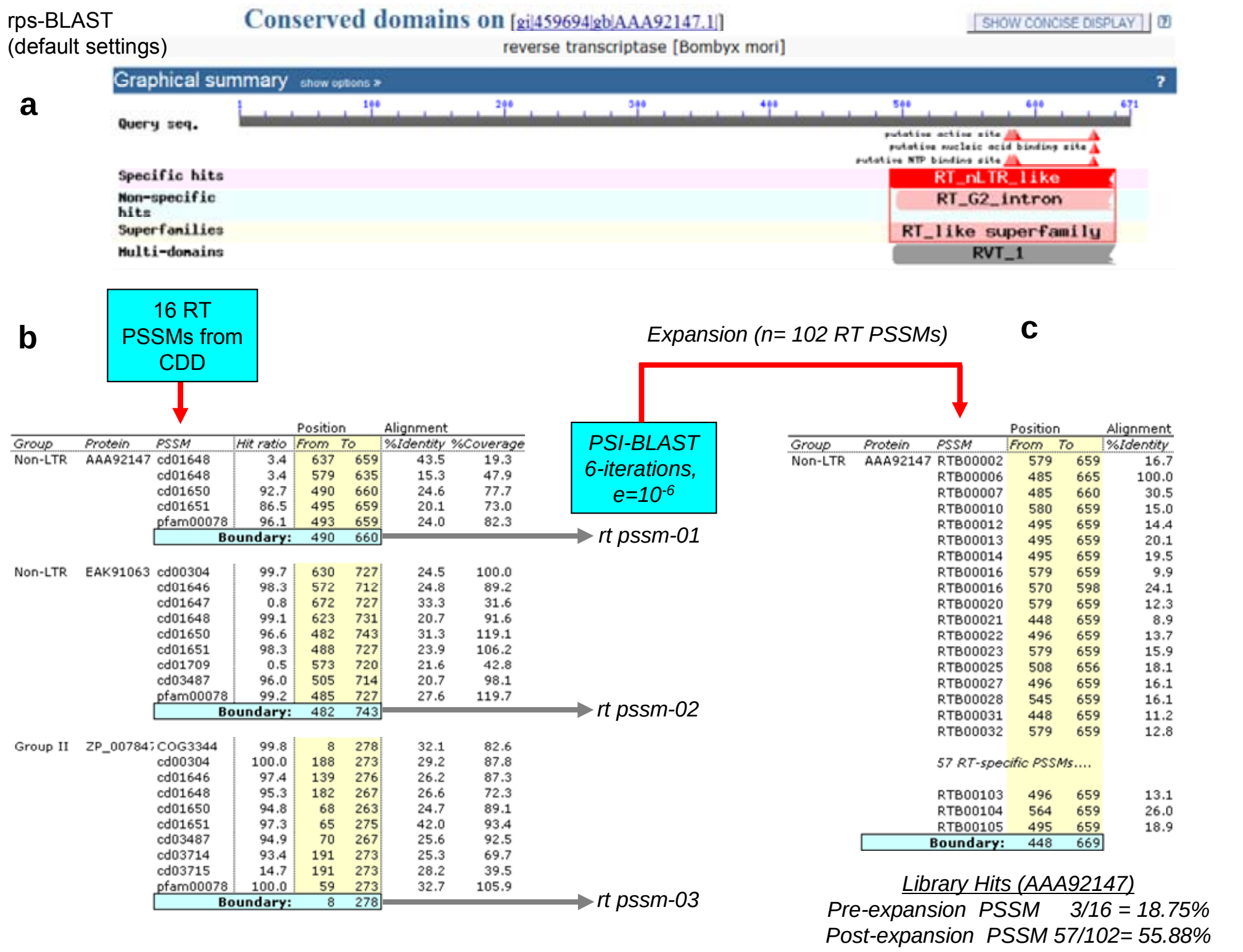

**Figure 2: Enrichment and Amplification of Signal Source PSSMs- (a)** Family/Superfamily specific PSSMs can be identified using NCBI Conserved Domain Database. (**b)** rpsBLAST can then be used to identify overlapping alignments between individual queries and family/superfamily specific CDD PSSMs. Overlapping alignments are then used to define domain specific region as described in methods. (**c)** Domain specific regions as identified are then used to generate PSSMs using PSI-BLAST and NCBI non-redundant (nr) database.

## Phylogenetic Trees generated using PHYRN

### *Phylogenetic Reconstruction of the RT superfamily*

Despite the low-identity and high variance in % gaps and coverage, when a phylogenetic tree is constructed from a phylogenetic profile comprised of 100 RT query sequences and 102 RT-specific PSSMs using Neighbor-Joining (11) (see Methods), we obtain a robust monophyletic tree with deep statistical support (bootstrap and jackknife, Figure 3c). In all aspects, this tree is superior to the tree constructed pre-expansion of the RT-specific PSSMs (Fig 3 in Chang et al (2)). Specifically, the Hepadnaviruses now form a clear monophyletic clade, and the Mt Plasmids now reside in the prokaryotic group as expected.

As previously mentioned, phylogenetic profile measurements improve with larger datasets (1;3;7). Therefore, we wondered whether our resolution could be increased by the inclusion of additional sequences across multiple taxa, thereby increasing the size of the PSSM library. We collected 716 full-length RT containing sequences from the literature (14-20) and PSI-blast aided searches of NCBI non-redundant database. These sequences were subsequently included in our RT-specific PSSM library as previously described. Figure 4 depicts a linearized phylogenetic tree of 716 full length retroelements measured using 846 PSSMs generated from the RT domain and rooted with retrointrons. The pairwise distances among them were acquired based on Euclidean distance measurement in the 716 $\times$ 846 data matrix, and an unrooted phylogenetic tree was derived from the 716 $\times$ 716 distance matrix using a neighbor-joining method. The tree is drawn to scale, with branch lengths in the same units as those of the Euclidean distances calculated from the data matrix.

Even at this size and level of divergence (~17% identity between groups), the PHYRN tree has robust monophyly. Within these monophyletic nodes, there are multiple subclades which are evident from this analysis. For example, we observe numerous subgroups of TY3 retroelements, and proper subclade groupings of retroviral RTs. Recently, Simon et al reported that there is a plethora of uncharacterized bacterial RTs (19). Our analysis is congruent with this proposal as we also observe novel clades of bacterial origin.

While these results are promising, the evolutionary history is unknown and therefore we cannot fully evaluate the performance of PHRYN using this dataset. Further, during the course of these experiments, we determined that although effective, the pipeline as described is computationally expensive for large datasets. To overcome these limitations, we made the following changes to the pipeline. Specifically, we compiled all PSSMs into a single database that can be used by standard

rps-BLAST(8). This drastically reduced the number of operations and enabled us to use a sliding window of e-value thresholds to recover positive alignments. In data not shown, these changes afforded us a >600-fold increase in computational speed as well as improved speciation (see Methods for complete details).

**a**

$$score = \%i \times \%c \text{ (\%i: Percent identity, \%c: Percent coverage)}$$
$$= \frac{ids}{alen} \times \frac{aqlen}{plen}$$
$$= \frac{ids}{plen} \times \frac{aqlen}{aqlen + gaps}$$

*ids*: The number of identical residues

*alen*: The alignment length including gaps

*aqlen*: The alignment length in the query excluding gaps ( $= query_to - query_from + 1$)

*plen*: The sequence length of a PSSM

*gaps*: The number of gaps in the query-sided alignment

$$\because alen = aqlen + gaps$$
$$\because \frac{aqlen}{aqlen + gaps} = \text{(Gap weight based on the number of gaps in the query-sided alignment)}$$

$$= \text{(Proportion of amino acid sites identical over a domain)} \times \text{(Gap weight)}$$
$$= \text{(1 − Proportion of amino acid sites different over a domain)} \times \text{(Gap weight)}$$
$$= (1 - p) \times w$$

**b**

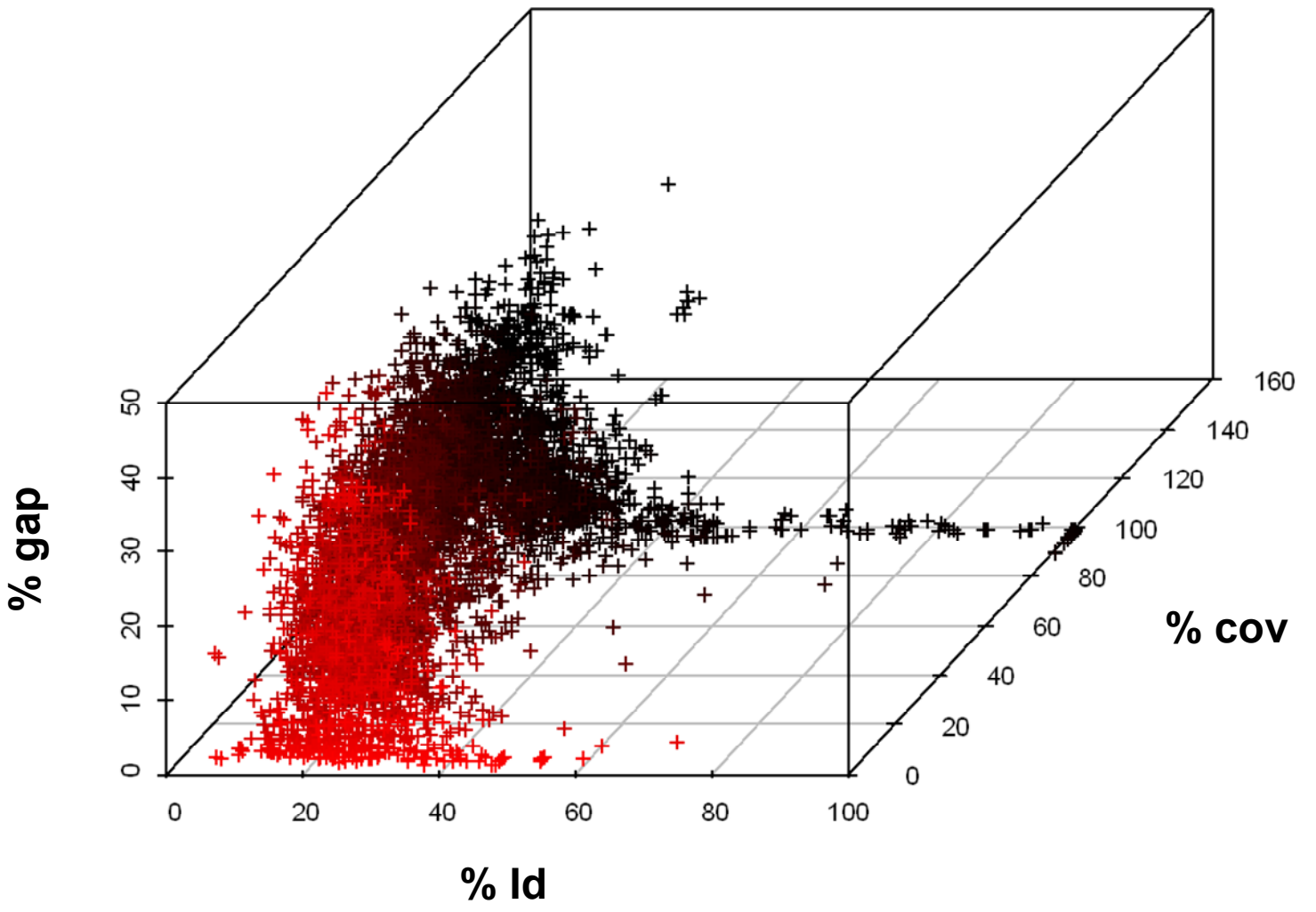

**c**

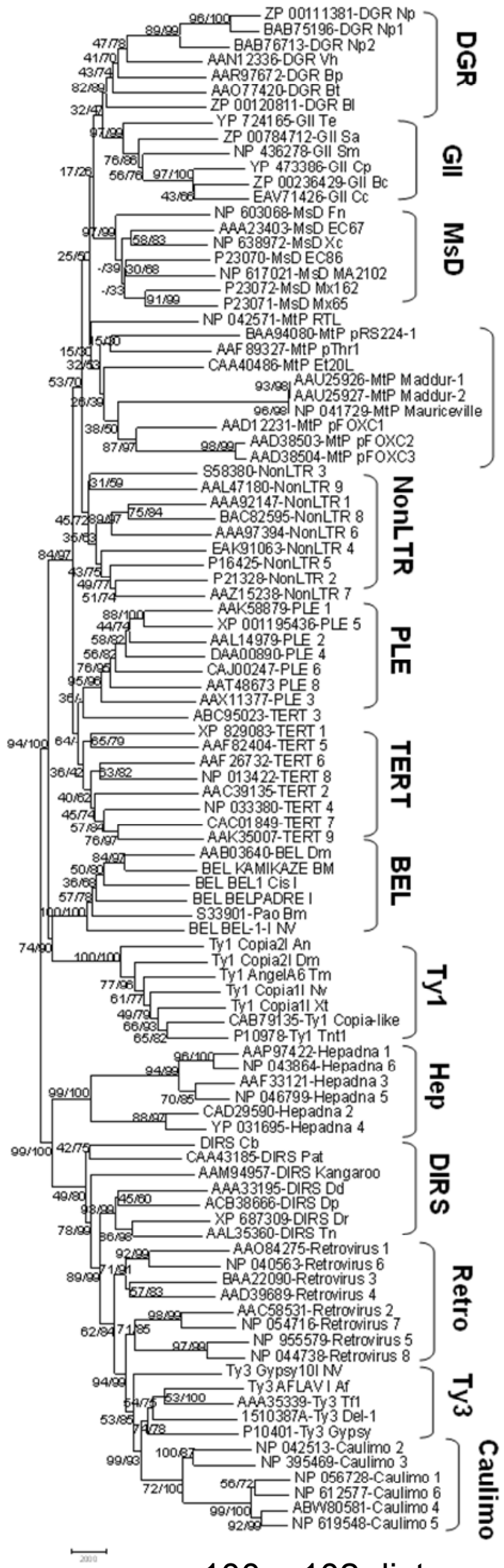

**Figure 3: Phylogenetic profile based measurements of evolutionary distance.**- **(a)** Algebraic derivation of p-distance from PHYRN product scoring scheme (%Identity x %coverage). **(b)** Distribution of 100 Retroelements by measurements of % Identity, % coverage, and %gap. **(c)** Unrooted phylogenetic tree of 100 full length retroelements measured using 102 PSSMs generated from the RT domain. The pairwise distances among them were acquired based on Euclidean distance measurement in the 100 X 102 data matrix, and an unrooted phylogenetic tree was derived from the 100 X 102 distance matrix using a minimum evolution method. The tree is drawn to scale, with branch lengths in the same units as those of the Euclidean distances calculated from the data matrix. Bootstrap and jackknife (80% fraction of samples) values were obtained from 1,000 replicates and are reported as percentages.

### *Phylogenetic Reconstruction of Rose Simulations*

Rose (Random Model of Sequence Evolution - Version 1.3; http://bibiserv.techfak.uni-bielefeld.de/rose/) implements a probabilistic model for protein sequence evolution (9). In this simulation, sequences are created from a common ancestor to produce a dataset of known size, divergence, and history. In this artificial evolutionary process, the accurate history is recorded since the multiple sequence alignment is created simultaneously. This allows us to have perfect control over evolutionary rates, allowing us to test the efficacy of our approach.

Figure 5 provides the results from PHRYN and MUSCLE (21) using 67 sequences simulated for 17% identity by ROSE (see Methods for full description of PSSM generation). Whereas MUSCLE performs poorly on this dataset (Figure 5b), PHYRN recaptures 93 % of the true evolutionary history and has only one deep node incorrectly identified (Figure 5a). In a second simulation, we maintained a similar level of divergence while increasing the size of the simulated dataset to 584 sequences. In this simulation, MUSCLE performance decreases significantly (no deep-nodes are correctly obtained, (Figure 6a), while PHRYN performance is still robust (Figure 6b). PHYRN recaptures 76% of the true evolutionary history at the deepest 64 nodes. Taken together, these results demonstrate the power of PHYRN for deriving deep evolutionary information.

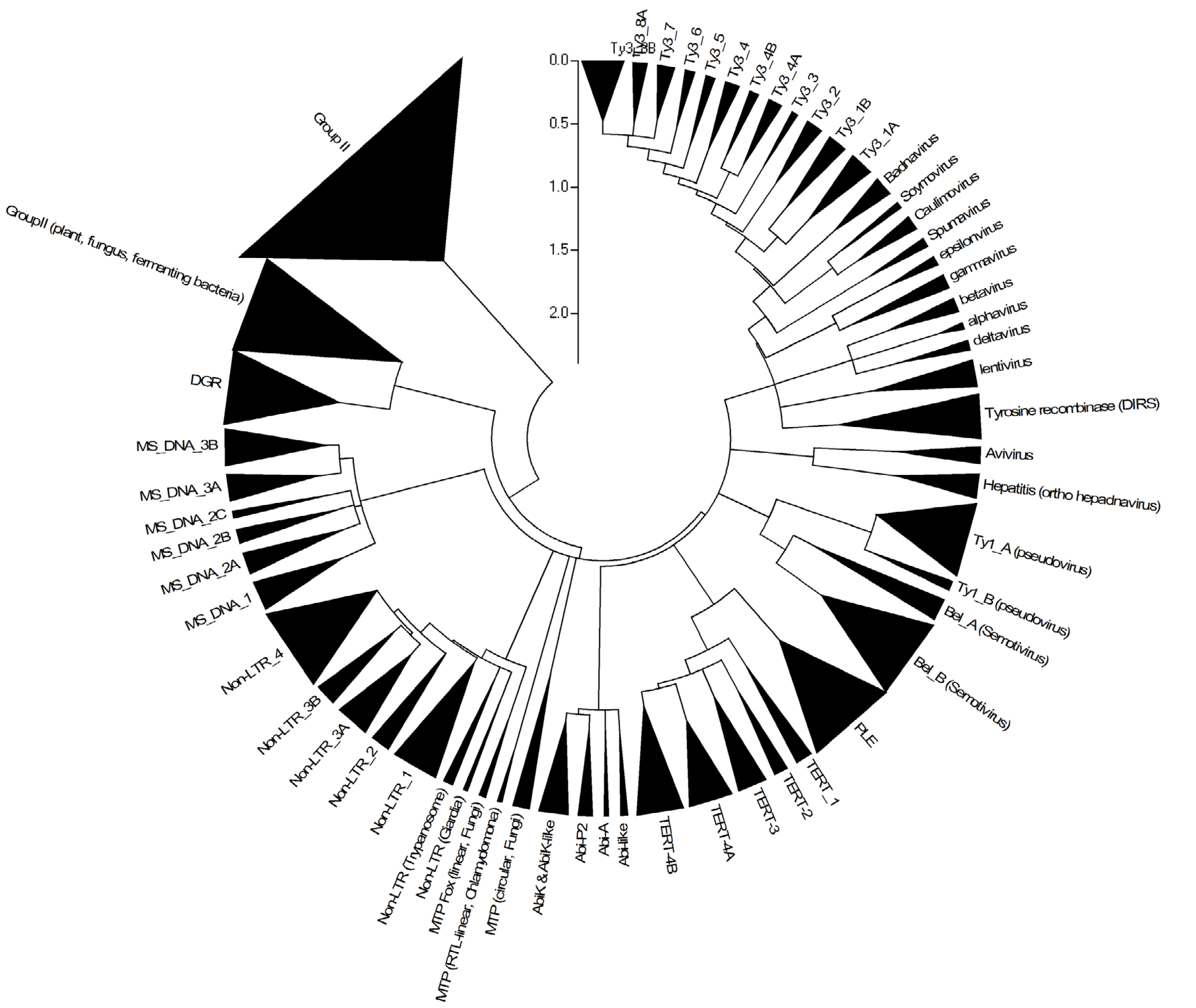

**Figure 4: Towards comprehensive phylogenies.-** Linearized phylogenetic tree of 716 full-length retroelements measured using 846 PSSMs generated from the RT domain and rooted with retrointrons. The pairwise distances among them were acquired based on Euclidean distance measurement in the 716 X 846 data matrix, and an unrooted phylogenetic tree was derived from the 716 X 716 distance matrix using a neighbor-joining (NJ) method. The tree is drawn to scale, with branch lengths in the same units as those of the Euclidean distances calculated from the data matrix.

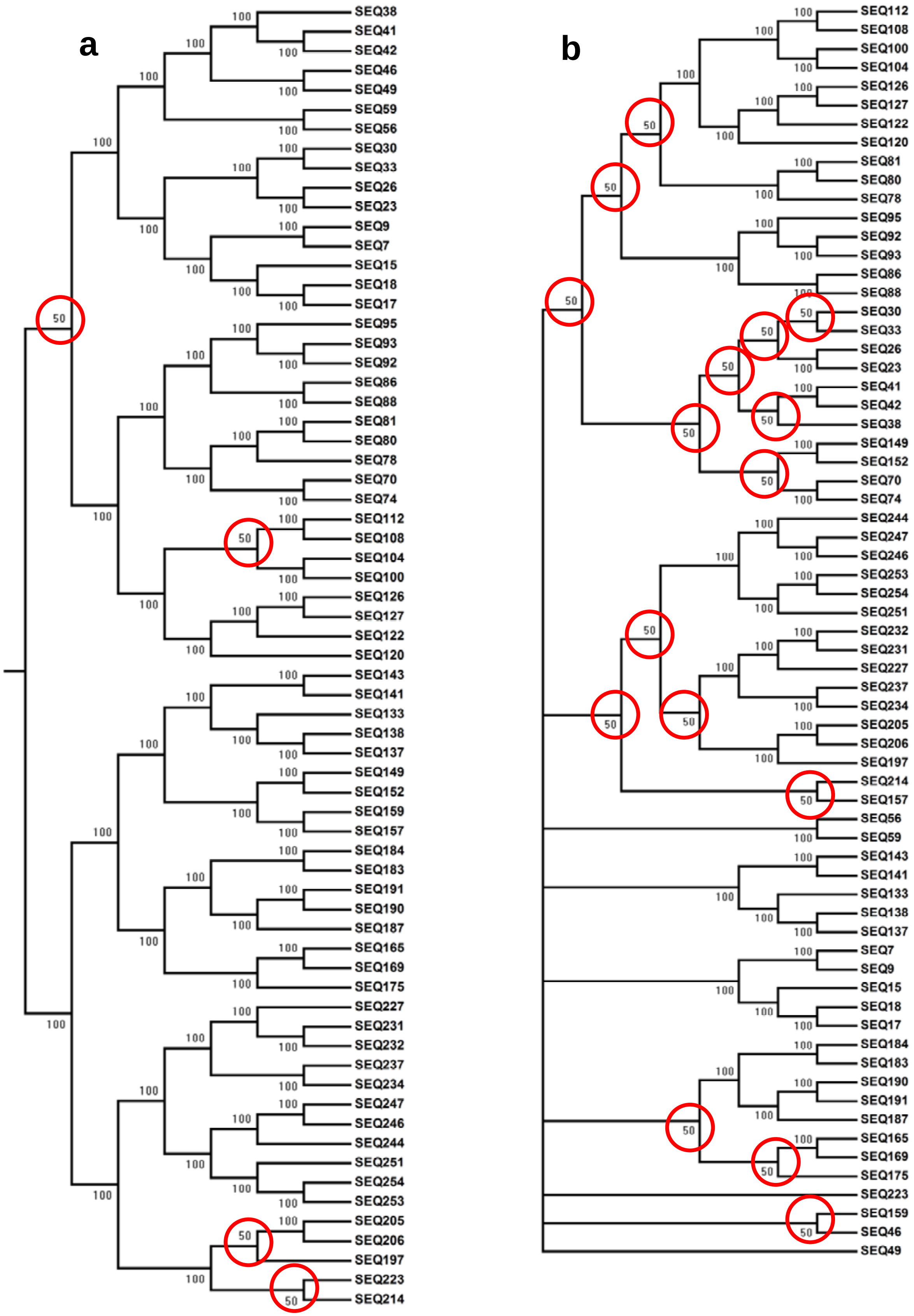

**Figure 5: PHYRN recapitulates 'true evolutionary history' better than MUSCLE in simulated protein families-** Consensus tree between original ROSE tree and tree generated using a) PHYRN and b) MUSCLE. Simulated protein family generated using ROSE, with an average distance of 550 (p distance ~0.83). Red circles mark the branch points (nodes) that are not recapitulated correctly. (no. of query sequences = 67)

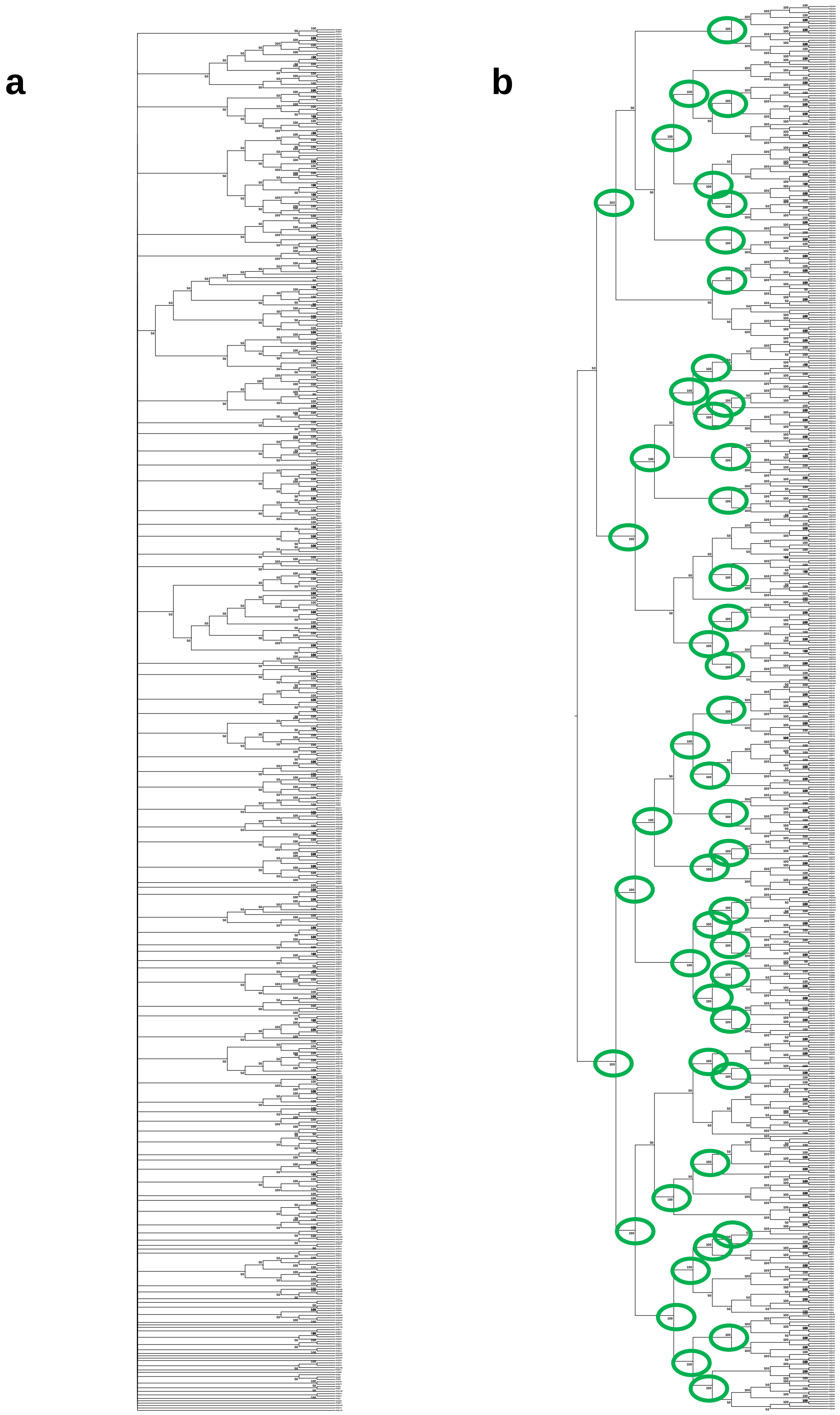

**Figure 6: Deep Node Recapitulation of 'true evolutionary history' in mega-phylogenies-** Consensus tree between original ROSE tree and tree generated using a) MUSCLE and b) PHYRN. Simulated protein family generated using ROSE, with an average distance of 550 (p distance ~0.82). Green circles mark the deep nodes that are recapitulated correctly in the consensus trees. (no. of query sequences = 584)

## Discussion

Our case-study of the RT superfamily and simulated datasets demonstrates that PHYRN is capable of inferring deep evolutionary relationships between highly divergent proteins. A number of implications can be derived from this study: (i) phylogenies built with PHYRN recapture more of the true evolutionary history and have robust statistical support; (ii) phylogenies built on pairwise alignments outperform conventional MSA methods and (iii) this method is scalable to thousands of sequences. This improved performance is due the improved information content contained in the PSSM libraries used in this study. We improved the efficacy of our PSSMs by: (i) limiting the PSSMs to homologous domains, (ii) optimizing the PSI-BLAST settings for their generation, and (iii) creating a pipeline that is sufficiently fast to handle large datasets.

Conversely, with respect to MSA dependent methods, increasing the number of query sequences makes it increasingly difficult to obtain an optimal multiple sequence alignment(22); in PHYRN, increasing number of query sequences also increases the dimensionality of the phylogenetic profile, thus increasing the alignment information space. This increase in information space leads to better, more robust measurements of relative rates. This 'comprehensive survey' approach, where more sequences are better, is in contrast to 'random walk' approach of MSA dependent methods where increased sequences are a problem and trees are limited to discrete taxa. Further, use of frequency tables in the phylogenetic profiles provides more informative measurements for calculating relative rates of evolution. This approach provides PHYRN with a potential to generate trees with thousands of sequences where the only theoretical limit is the available sequencing data. Indeed, when we expanded the RT tree from 100 to 716 sequences comprising >14 groups we obtain a tree that is consistent yet higher resolution than previously reported RT studies (2;15-17).

As PHRYN is well suited to making measurements on large divergent datasets, we hypothesize this approach may be capable of solving a number of unanswered questions related to the ancient origins of life and speciation. Moreover, since PHYRN functions in the twilight zone of sequence similarity, this algorithm may have the ability to inform whether functionally or structurally similar proteins have a common ancestor or occurred via convergent evolution. In conclusion, our study provides strong evidence that, even in its nascent stage, PHYRN measurements can provide key insight into evolutionary relationships among distantly related and/or rapidly-evolving proteins.

**Acknowledgements-** This work was supported by the Searle Young Investigators Award and start-up money from PSU (RLP), NCSA grant TG-MCB070027N (RLP, DVR), The National Science Foundation 428-15 691M (RLP, DVR), and The National Institutes of Health R01 GM087410-01 (RLP, DVR). This project was also funded by a Fellowship from the Eberly College of Sciences and the Huck Institutes of the Life Sciences (DVR) and a grant with the Pennsylvania Department of Health using Tobacco Settlement Funds (DVR). The Department of Health specifically disclaims responsibility for any analyses, interpretations or conclusions. We would like to thank Teresa Killick, Sree Chintapalli, and Anand Padmanbha for their help and support during the project as well as Jason Holmes at the Pennsylvania State University CAC center for technical assistance. We would also like to thank Drs. Robert E. Rothe, Jim White, Glenn M. Sharer, Cookie van Volmar, Barbara VanRossum, Russell Hamilton Carroll, C. Heesch and C. Hong for creative dialogue.

M (no. of PSSMs) matrix of product scores. Euclidean distance was calculated based on the PHYRN product score of each query and an N X N Euclidean distance matrix was generated. Similar to method used for RT trees, phylogenetic trees were inferred using Neighbor Joining (NJ) and Minimum Evolution (ME) method, as available in MEGA (13).

**Generating phylogenetic trees using MUSCLE**

Optimal Multiple Sequence Alignment (MSA) for a given dataset was obtained using MUSCLE v3.6 (23). Phylogenetic trees for these optimal MSA were inferred using MEGA's Neighbor-joining (NJ) and Minimum-evolution (ME) algorithm, with pairwise deletion and p-distance as default settings.

**Consensus trees with ROSE 'true history'**

We used 'consense' program of PHYLIP v3.67 package (24;25) to generate consensus trees between PHYRN and Rose trees, as well as between MUSCLE and Rose. Recapitulation rate and percentages were then calculated from consensus tree newick files.

**Bootstrap and Jackknife**

We generated 3,000 random samples from our PHYRN M-dimension, using random number generator code from PHYLIP source code (http://evolution.genetics.washington.edu/phylip.html). During resampling, same columns were allowed to be selected more than once. We then used Fitch program with default settings in PHYLIP 3.67 package to generate minimum-evolution (ME) trees for each sample, followed by Consense program for generating a consensus trees of all samples by majority rule. For jackknife resampling, we followed a similar approach to generate 1,000 random samples, however only 80% of original M-dimensional data was resampled each time. FITCH and Consense programs were then used in similar manner as used in bootstrap resampling.